\documentclass[aps,prb,10pt,amsmath,amssymb]{revtex4}
\usepackage{amssymb}
\usepackage{amsbsy}             
\usepackage{bm}
\usepackage{mathrsfs}

\usepackage{dcolumn}
\usepackage{amsmath}
\usepackage{epsfig}

\usepackage{latexsym}

\begin{document}
                                                                                                                         
\title{Path integral formulation for quantum nonadiabatic dynamics and the mixed 
quantum-classical limit.}

\author{Vinod Krishna\footnote{Present address, Dept of Chemistry, University of Utah, 315 So, 1400 E.
Salt Lake City, UT 84102. email:vkrishna@hec.utah.edu}} 
\affiliation{Department of Physics, Yale University, New Haven, CT-06520}
\begin{abstract}  
   This work identifies geometric effects on dynamics due to nonadiabatic
couplings in Born Oppenheimer systems and provides a systematic method for deriving
corrections to mixed quantum-classical methods. Specifically, an exact path integral 
formulation of the quantum nonadiabatic dynamics of Born Oppenheimer systems is described. 
Stationary phase approximations to the propagator for full quantum dynamics are derived.
It is shown that quantum corrections to mixed quantum classical methods can be obtained 
through stationary phase approximations to the full quantum dynamics. A rigorous 
description of the quantum corrections due to electronic nonadiabatic coupling on the 
nuclear dynamics within the Ehrenfest framework is obtained. The fewest switches surface 
hopping method is shown to be obtained as a quasiclassical approximation to the dynamics 
and natural semiclassical extensions to include classically forbidden nonadiabatic 
transitions are suggested. 
\end{abstract}
\maketitle

 


\pagenumbering{arabic}
\baselineskip 13pt
\baselineskip 17pt
 

 
 

\section*{I. Introduction}
     The Born-Oppenheimer approximation$\cite{BornO}$ is fundamental to studies of a 
wide variety of molecular systems. However, violations of the Born-Oppenheimer 
approximation$\cite{Butler}$ are ubiquitous in many molecular and condensed phase phenomena, 
and are responsible for several relaxation processes. Despite their importance, fully quantum 
treatments of nonadiabatic corrections to Born Oppenheimer dynamics are limited and no perturbative
formulation exists to account for them.  
     
     It is practically impossible to compute the full quantum nonadiabatic dynamics for realistic 
systems due to the prohibitive computational cost associated with such a calculation. To circumvent 
this difficulty, a class of mixed quantum classical methods have been invented$\cite{Ehren,pech1,pech2,
tully1,tully2,Kapral,Ohrn,Ohrn1}$. These methods can be broadly classified into 
mean-field$\cite{Ehren,Ohrn,Ohrn1}$ and trajectory surface hopping methods$\cite{tully1,tully2}$. 
Mean field methods originate from work by Ehrenfest$\cite{Ehren}$, where the nuclear 
degrees of freedom, treated as classical mechanical variables, experience a potential that 
is the expectation value of the electronic Hamiltonian with respect to the instantaneous electronic 
wavefunctions obtained by solving the electronic Schrodinger equation. On the other hand, surface 
hopping methods$\cite{tully1,tully2}$ treat the nuclear dynamics as being localized to a single 
Born-Oppenheimer surface at any time, with nonadiabatic couplings between surfaces leading to 
instantaneous jumps, or "hops" between Born-Oppenheimer surfaces. Common to all mixed quantum-classical 
methods is the identification of the nuclear (or slow) degrees of freedom as behaving as essentially 
classical quantities, while interacting with a quantum system consisting of the electronic degrees 
of freedom. This identification is usually justified by a mixture of mathematical and phenomenological 
arguments. As a result, while these methods have been investigated by intuitive tests and computational 
simulations over the years, there is no rigorous theory to understand situations where this separation 
into classical and quantum subsystems breaks down and quantum effects on the classical degrees of
freedom have to be considered. This problem is also evident in studies of the dynamics of 
molecular systems on higher-dimensional Born-Oppenheimer surfaces, where geometric features, like
conical intersections between surfaces, have a significant physical effect on the dynamics.    
Thus, given the somewhat ad-hoc nature of most mixed quantum classical methods, a central problem is
to develop a framework which would incorporate these methods as an approximation and would enable
the systematic study of corrections to the mixed quantum classical picture of nonadiabatic dynamics.
     
This work aims to systematically construct mixed quantum classical methods as semiclassical limits 
of quantum nonadiabatic dynamics. Such an approach enables the systematic improvement of mixed quantum 
classical methods to include further quantum corrections as well as the development of new and hybrid 
mixed quantum classical methods. Exact path integral descriptions of the quantum 
nonadiabatic dynamics are provided and stationary phase approximations for the dynamics are derived. 
The resulting dynamical picture is a consistent means of obtaining a mixed quantum classical approximation 
to the nonadiabatic dynamics. It is demonstrated that both the Ehrenfest approach and the surface 
hopping approach correspond to stationary phase approximations of the quantum nonadiabatic dynamics, 
when studied in complementary representations of the electronic system. Furthermore, the approximate 
equations of motion obtained in the Ehrenfest picture contain additional contributions to the effective 
nuclear force. These contributions are a direct, and nontrivial consequence of the geometry of the 
Born-Oppenheimer potential energy surfaces. For systems with only one nuclear dimension, these 
contributions are zero, and the dynamics is of pure Ehrenfest type, while for higher dimensional 
surfaces, they are non-zero, and represent corrections to the quasiclassical dynamics due to the 
topological features of the surface, like the presence of conical intersections and degeneracies. 
In the complementary picture from which surface hopping methods are obtained as an approximation, 
quantum corrections to the nonadiabatic nuclear dynamics can be built in by considering semiclassical 
expansions of the nuclear propagator in the Born-Oppenheimer approximation. In addition to this, 
the approach outlined in this paper provides a new interpretation of nonadiabatic effects as 
a consequence of the geometry of the electronic Hilbert space, and thus as arising due to a 
generalization of Berry's phase for parametrized quantum systems. 

The outline of the paper is as follows. In Sec.II, a path integral description of the quantum 
nonadiabatic dynamics is introduced. Issues concerning path integral approaches to nonadiabatic 
dynamics are discussed and the derivation of the approach used in this paper are outlined in 
Sec.II.A. In Sec.II.B, the path integral description is derived. It is based on integrating out
the electronic degrees of freedom using a coherent state basis. Furthermore, the path integral 
so obtained is compared to other path integral formulations constructed within a Born-Oppenheimer
basis. It is shown in this section that nonadiabatic couplings between states are due to the 
geometry of the electronic Hilbert space, and can be regarded as generalizations of Berry's 
phase. In Sec.III, the path integral approach is analysed and stationary phase approximations 
are derived. Sec.III.A demonstrates that the stationary phase approximation to the coherent
state path integral corresponds to the Ehrenfest formulation of mixed quantum classical 
dynamics. Furthermore, geometric corrections to the nuclear force in the Ehrenfest approach
are obtained. These corrections are related to the phase coherence between different states, 
and are most significant when different Born Oppenheimer energy surfaces become close in energy and 
are strongly coupled due to nonadiabatic effects. Sec.III.B studies stationary phase approximations 
to the complementary path integral description constructed using a Born-Oppenheimer basis for 
the electronic system. It is shown that the equations of motion so obtained correspond in the
semiclassical limit, to the fewest switches surface hopping approach to nonadiabatic dynamics. 
The theory in Sec.III.B also points to the construction of semiclassical propagators that can 
extend descriptions of nonadiabatic dynamics beyond the mixed quantum-classical limit. 
Sec.III.C describes semiclassical quantization rules to approximate the eigenvalue 
spectrum for the full quantum dynamics. The paper then concludes with a discussion of 
the results and future work in Sec.IV.  
   

\section*{II. Theory}
  The problems with path integral approaches to quantum nonadiabatic dynamics are 
briefly discussed, and a short outline of the consequent derivation of the path 
integral based theory is given. The detailed mathematical treatment of a path 
integral approach based on coherent states is then subsequently discussed. 

  It should also be noted that this text uses the units convention where
Planck's constant is set to one, i.e  $\hbar =1$ in this paper. 
\subsection{Preliminaries}
  The nonadiabatic corrections to the Born Oppenheimer approximation can be
viewed as a result of a coupling between a set of "fast" or bath degrees of freedom and
slow or "system" degrees of freedom. However, the mechanisms that result from this
coupling are different from the usual problem of a system interacting with a dissipative
bath, since the Hilbert space enclosing the bath degrees of freedom in the Born Oppenheimer
theory is parametrized by the slow degrees of freedom. As a consequence, the time dependent
evolution of such a system involves complex geometric phase effects$\cite{berry,zee,mead1,
zygelman1,shapere1,Iida}$. Pechukas$\cite{pech1,pech2}$ first developed a path integral 
prescription for nonadiabatic dynamics and derived a set of nonlocal equations of motion for 
the semiclassical nuclear path. Later work$\cite{shapere2,Iida}$ on a path integral description 
clarified the interpretation of the nonadiabatic coupling vector potentials as nonabelian 
gauge fields. However, these descriptions are not easily amenable to approximations which can 
be studied computationally. In the Pechukas formulation, the effect of the electronic
degrees of freedom on system dynamics enters through a nonlocal potential which is a
function of the paths taken by the system, and similarly in the formulation of Moody,
Shapere and Wilczek$\cite{shapere1,shapere2}$, nonadiabatic couplings between Born-Oppenheimer
states enter as path ordered matrix exponentials. Thus, an approach that adopts the path
integral framework for the exact nonadiabatic dynamics, but which is also amenable to
systematic and computationally tractable approximations would be of considerable theoretical
and practical utility.

      The problems with regards to nonlocality and path ordered exponentials encountered in a
path integral description of the quantum nonadiabatic dynamics can be partially resolved by an
appropriate choice of the path integral. To this end, a local basis of continuous, $N$-particle 
coherent state Slater determinants is chosen to describe the electronic system. The use of coherent
states enables the straightforward construction of stationary phase approximations to the quantum
dynamics and corresponding semiclassical quantization rules. Furthermore, since the propagator 
can be written as a functional integral over a continuous coherent state space, issues relating
to path ordering and nonlocality become considerably simplified. 
     
      The coherent state basis is parametrized by the nuclear coordinates and is overcomplete. 
Since the basis is parametrized by the nuclear coordinates, a local 
Hilbert space is defined at each point on a given nuclear path. Thus, intuitively, motion 
along a nuclear path involves the following: The first contribution is due to the propagation of 
the nuclear part of the Hamiltonian, and the second is due to a rotation of the Hilbert space 
from one point into the next one on the path (Fig 1). This rotation takes basis states at one point 
into the basis states at the next, and hence changes the electronic contribution to the effective 
nuclear Hamiltonian. In addition, the rotation of the basis in general involves "twists" and skews 
resulting in torsional contributions to the force acting on the nuclear degrees of freedom. 

     To explicitly describe this, the derivation of the coherent state path integral proceeds 
as follows. The matrix elements of the time propagation operator for the full (nuclear plus 
electronic) Hamiltonian between initial and final nuclear positions are written down as a path 
integral. The time for propagation is broken up into a large number of (infinitesimal) 
time segments and the propagation operator is written in a Trotter product expansion. 
Complete sets of nuclear position eigenstates are then introduced for each time segment, 
and an infinitesimal propagator between successive nuclear positions is written down. The 
resulting expression is a path integral of an operator acting on the electronic 
degrees of freedom over each nuclear path. For a given nuclear path, this quantity simply 
corresponds to the electronic propagation operator parametrized by the nuclear path. 
     
   The matrix elements of the quantum nonadiabatic propagation operator between initial and 
final electronic states and nuclear positions are then evaluated through the matrix
elements of this path dependent quantity between initial and final electronic states. 
This is done by adding overcomplete sets of electronic $N$ particle coherent state Slater 
determinants at each time slice and evaluating matrix elements of the infinitesimal 
propagation operator. The overlap between these basis states belonging to consecutive 
timesteps gives rise to the nonadiabatic coupling contribution to the dynamics. The 
resulting path integral is then further analysed.    
\subsection{Derivation}
   The full Hamiltonian for a system with nuclear and electronic degrees of freedom can be 
written as
\begin{equation}
   H = H_{n}({\bf R}) + H_{e}({\bf r, R}) 
\end{equation}
   The nuclear Hamiltonian $H_{n}(\bf{R})$ is given by
\begin{equation}
    H_{n}({\bf R}) = \sum_{I=1}^{K}{\frac{{\bf P}_{I}^{2}}{2M_{I}}} + U({\bf R})
\end{equation}
   The coordinates ${\bf R}=(\vec{R}_{1},...\vec{R}_{K})$ denote the $K$ nuclear degrees of 
freedom with momenta given by ${\bf P} = ({\bf P}_{1},{\bf P}_{2},...{\bf P}_{K})$. The potential energy
function, $U({\bf R})$ is the bare nuclear potential. The electronic Hamiltonian can be similarly 
defined and the electronic coordinates, $\{\vec{r}_{1},\vec{r}_{2}...\vec{r}_{N}\}$, are elements 
of the vector $\bf r$. The electronic potential energy $V({\bf r},{\bf R})$ is parametrically 
dependent on the nuclear positions.
\begin{equation}
    H_{e}({\bf r, R}) = \sum_{i=1}^{N}{\frac{{\bf p}_{i}^{2}}{2m_{i}}} + V({\bf r, R})
\end{equation}
   For simplicity of presentation, the nuclei and electrons are assumed to be spinless. This 
assumption is reasonable for the nonadiabatic effects considered here, and the treatment that 
follows can be extended to include spin effects. Also it is assumed that the nuclear masses,
$M_{I}$ are all equal and have the value $M$. This is solely for notational convenience and 
does not significantly affect the conclusions of this work. 
 
   A Born Oppenheimer basis of eigenstates and eigenvalues is defined by
\begin{equation}
\big[H_{e}({\bf r, R})\big]\mid\Psi_{\mu};{\bf R}\rangle = E_{\mu}({\bf{R}})\mid\Psi_{\mu};
{\bf{R}}\rangle
\end{equation}  
   The unitary time propagation operator for this Hamiltonian can be expanded in a 
Trotter product expansion as described below.
\begin{eqnarray}
e^{-iHt} = \prod_{k=1}^{P}{\exp{\Big[-i\frac{\epsilon}{2}(H_{e} + U)\Big]}
\exp{\Big[-i\epsilon\frac{{\bf P}^{2}}{2M}\Big]}\exp{\Big[-i\frac{\epsilon}{2}(H_{e} + U)\Big]}}
\end{eqnarray}
  The propagation operator for a given set of initial and final nuclear positions $\bf Q$ and 
$\bf Q'$ can be written as
\begin{eqnarray}
\langle{\bf Q}'\mid e^{-iHt}\mid{\bf Q}\rangle = 
\int{\prod_{k=1}^{P}{d{\bf R}_{k}\exp{\Big[-\frac{i\epsilon}{2}\{H_{e}({\bf r,R}_{k})+
U({\bf R}_{k})\}\Big]}\langle{\bf R}_{k}\mid\exp{\Big[-i\epsilon\frac{{\bf P}^{2}}{2M}\Big]}
\mid{\bf R}_{k+1}\rangle}}\nonumber\\
\exp{\Big[-\frac{i\epsilon}{2}\{H_{e}({\bf r,R}_{k+1})+U({\bf R}_{k+1})\}\Big]} & &
\end{eqnarray}
    where ${\bf R}_{1} = {\bf Q}$ and ${\bf R}_{P} = {\bf Q}'$. 
   It is important to note that the matrix elements of the propagation operator described in
Eq.(6) are still operators acting on the electronic Hilbert space. This is because the matrix 
elements are a partial trace with respect to the eigenstates $\mid{\bf Q}'\rangle$ and 
$\mid{\bf Q}\rangle$ of the nuclear position operator $\hat{\bf R}$. 
   This partial trace can be rewritten as a path integral over the set of nuclear paths 
${\bf R}(t)$. 
\begin{equation}
\langle{\bf Q'}\mid e^{-iHt}\mid{\bf Q}\rangle = \int{D{\bf R}(\tau) e^{-iS_{n}[{\bf R}(\tau)]}
G_{na}[{\bf R}(\tau)]}
\end{equation}
   $S_{n}[{\bf R}(t)]$ is the bare nuclear action corresponding to the potential $U({\bf R})$.
   The quantity $G_{na}$ is an operator quantity in the electronic degrees of freedom, and
is a nonlocal functional of the nuclear paths. Physically, the operator $G_{na}[{\bf R}(\tau)]$ 
includes the "connection" between electronic Hilbert spaces defined at each point on a given 
nuclear path ${\bf R}(\tau)$. Thus, a complete solution of the nonadiabatic problem would require 
that electronic matrix elements of $G_{na}$ be evaluated. To perform such an evaluation, 
overcomplete sets of antisymmetric $N$ electron Slater determinant coherent state wavefunctions 
are introduced at each nuclear position along a given nuclear path$\cite{blaziot,NegOr}$. 

   A general $N$ electron antisymmetric wavefunction can be written as a linear combination of 
Slater determinants. This basis of $N$ electron Slater determinants, 
$\mid\phi\rangle=\mid\phi_{1},...\phi_{N}\rangle$ is overcomplete and has the closure property:
\begin{equation}
 1 = \int{\frac{d\phi^{*}d\phi}{(2\pi i)^{N}}e^{-\mid\phi\mid^{2}}\mid\phi\rangle\langle\phi\mid}
\end{equation}
 
   Here, a local basis of $N$ electron coherent state Slater determinants$\cite{blaziot,NegOr}$ 
is defined at each point on a given nuclear path. Hence, this local electronic basis is parametrized 
by the nuclear position to account for the fact that a local electronic Hilbert space exists at 
each nuclear position. For such a basis defined for a nuclear position $\bf R$, the completeness 
relation reads:
\begin{equation}
 1 = \int{\frac{d\phi^{*}({\bf R})d\phi({\bf R})}{(2\pi i)^{N}}e^{-\mid\phi({\bf R})\mid^{2}}\mid
\phi({\bf R})\rangle\langle\phi({\bf R})\mid}
\end{equation}
For notational convenience, the integration measure in the overcompleteness relation will be 
written as 
\begin{equation}
  d\mu[\phi({\bf R})] = \frac{d\phi^{*}({\bf R})d\phi({\bf R})}{(2\pi i)^{N}}e^{-\mid\phi({\bf R})\mid^{2}}
\end{equation}
  with the differentials and the absolute value of $\phi({\bf R})$ defined by
\begin{equation}
  d\phi({\bf R}) = \prod_{i=1}^{N}{d\phi_{i}({\bf R})};\quad \mid\phi({\bf R})\mid^{2} = \sum_{i=1}^{N}{\mid\phi_{i}({\bf R})\mid^{2}}
\end{equation} 
The parametrization of the electronic Hilbert space by the nuclear position is responsible
for geometric phase effects. As will be shown below, these effects enter through the definition 
of a local basis of wavefunctions at each nuclear position. 

The operator $G_{na}$ can be explicitly defined to be 
\begin{equation}
G_{na}[{\bf R}(t)]=\prod_{k=1}^{P}{\exp{[-i\epsilon H_{e}({\bf r, R}_{k})]}}
\end{equation}
This operator is a path ordered quantity because the electron Hamiltonian does not 
commute with itself when evaluated at different points on a nuclear path. To evaluate 
the matrix elements of this operator, overcomplete sets of $N$-electron Slater determinant 
wavefunctions are introduced at each timestep. On doing so, the matrix elements of $G_{na}$
between an initial electronic state $\mid\Phi^{0};{\bf Q}\rangle$ and a final state 
$\mid\Phi^{f};{\bf Q}'\rangle$ can be written as 
\begin{eqnarray}
\langle\Phi^{0};{\bf Q'}\mid G_{na}\mid\Phi^{f};{\bf Q}\rangle = 
\int{\prod_{j=1}^{P}{d\mu[\psi_{j}({\bf R}_{j})]d\mu[\phi_{j}({\bf R}_{j})]\langle\Phi^{0};{\bf Q'}
\mid\psi({\bf Q}')\rangle\langle\psi({\bf Q}')\mid e^{-i\epsilon H_{e}({\bf r,Q}')}
\mid\phi({\bf Q}')\rangle}}\nonumber\\
\langle\phi({\bf Q}')\mid\psi_{1}({\bf R}_{1})\rangle\langle\psi_{1}({\bf R}_{1})\mid 
e^{-i\epsilon H_{e}(\bf{r, R}_{1})}\mid\phi_{1}({\bf R}_{1})\rangle ...
\langle\phi_{P}({\bf R}_{P})\mid\Phi^{f};{\bf Q}\rangle\nonumber\\
\end{eqnarray}
  Two sets of overcomplete basis states are introduced at each timestep, because the time 
evolution operator is in general not diagonal in this basis of states, and the overlap 
between basis states corresponding to different positions on the nuclear path is not unity. 
More specifically, the coherent states inserted at successive timesteps, $t_{k},t_{k+1}$,
are parametrized by different sets of nuclear positions, ${\bf R}_{k},{\bf R}_{k+1}$.
Since there is no constraint on the basis states at different nuclear positions, all possible
matrix elements of the time evolution operator between these basis states will have  
to be considered. Furthermore, to evaluate this off-diagonal matrix element, an additional 
set of basis states parametrized by the position ${\bf R}_{k}$ is 
introduced at time $t_{k+1}$. Since the path integral is discontinuous in the coherent 
state space, all possible overlaps of this new basis element with the basis element 
parametrized by ${\bf R}_{k+1}$ will have to be considered, and the matrix elements 
of the time evolution operator are not constrained to be diagonal. This lack of 
constraint is unlike the cases encountered with more traditional coherent state 
path integrals where only diagonal matrix elements between coherent states suffice, 
and is a consequence of the parametrization by the nuclear coordinates. 
 
It is clear that the above path integral involves matrix elements of the form
\begin{equation}
\Lambda[\psi({\bf R}_{k}),\phi({\bf R}_{k})] = \langle\psi({\bf R}_{k})\mid\exp{(-i\epsilon
H({\bf r,R}_{k}))}\mid\phi({\bf R}_{k})\rangle
\end{equation}
 If the exponent on the right hand side of Eq.(14) is expanded in powers of $\epsilon$ and terms 
higher than the first order are neglected, the matrix elements $\Lambda$ can be evaluated to be 
approximately
\begin{equation}
\Lambda[\psi({\bf R}_{k}),\phi({\bf R}_{k})] = 
\langle\psi({\bf R}_{k})\mid\Big[1 - i\epsilon H({\bf r,R}_{k})\Big]\mid\phi({\bf R}_{k})\rangle
\end{equation}
Making the same assumption as before, the right hand side of Eq.(15) can be reexponentiated to
yield
\begin{equation}
\Lambda[\psi({\bf R}_{k}),\phi({\bf R}_{k})] = 
\langle\psi({\bf R}_{k})\mid\phi({\bf R}_{k})\rangle\exp{\Big[-i\epsilon\frac{\langle
\psi({\bf R}_{k})\mid H({\bf r, R}_{k})\mid\phi({\bf R}_{k})\rangle}{\langle\psi({\bf R}_{k})
\mid\phi({\bf R}_{k})\rangle}\Big]}
\end{equation}
  In addition to the matrix elements $\Lambda$, the path integral contains 
"connection" factors of the form
\begin{equation}
\Gamma_{k} = \langle\phi({\bf R}_{k-1})\mid\psi({\bf R}_{k})\rangle
\end{equation}
  These connection factors are geometric in character and are due to the 
parallel transport of the local electronic basis from one nuclear position
to the next. Thus, these factors correspond to a generalization of the 
canonical, or Berry's phase for simple, parameter dependent Hamiltonians$\cite{berry}$. 
To obtain a more familiar form for this phase dependence, the connection 
factors can be rewritten as follows:
\begin{equation}
\Gamma_{k} = (1-\frac{\langle\delta\phi({\bf R}_{k})\mid\psi({\bf R}_{k})\rangle}{
\langle\phi({\bf R}_{k})\mid\psi({\bf R}_{k})\rangle})\langle\phi({\bf R}_{k})\mid
\psi({\bf R}_{k})\rangle  
\end{equation}
  The difference $\delta\phi({\bf R}_{k})$ is the difference between the wavefunctions $\phi$  
evaluated at successive time steps.
\begin{equation}
   \delta\phi({\bf R}_{k}) = \phi({\bf R}_{k}) - \phi({\bf R}_{k-1})
\end{equation}
 
   On exponentiating and making the assumption that the basis wavefunctions $\phi$ vary 
continuously, the connection factor takes the form 
\begin{equation}
\Gamma_{k} = \langle\phi({\bf R}_{k})\mid\psi({\bf R}_{k})\rangle\exp{\{i\epsilon\langle i\partial_{t}
\phi({\bf R}_{k})\mid\psi({\bf R}_{k})\rangle\}} + O(\epsilon^{2})
\end{equation}
where the time derivative acts on the wavefunction $\phi$. 

  Thus, on multiplying the matrix elements and correction factors together, the following 
path integral expression for the electronic contribution $G_{na}$ can be obtained as 
below:
\begin{eqnarray}
\langle\Psi^{0};{\bf Q}'\mid G_{na}\mid\Psi^{f};{\bf Q}\rangle = \int{D[\psi^{*}({\bf R}(\tau)),
\psi({\bf R}(\tau))]D[\phi^{*}({\bf R}(\tau)),\phi({\bf R}(\tau))]
\langle\Psi^{0};{\bf Q}'\mid\psi({\bf Q}')\rangle}\nonumber\\
\langle\phi({\bf Q})\mid\Psi^{f};{\bf Q}\rangle\exp{\Big[i\int_{0}^{t}
{\frac{\langle\phi({\bf R}_{\tau})\mid i\overleftrightarrow{\partial}_{\tau} - H_{e}
\mid\psi({\bf R}_{\tau})\rangle}{\langle\phi({\bf R}_{\tau})\mid\psi({\bf R}_{\tau})\rangle} d\tau}\Big]}
\end{eqnarray}
  The quantities $D[\psi^{*},\psi] = \prod_{j=1}^{P}{d\mu[\psi_{j}({\bf R}_{j})]}$ are
the measures for the coherent state path integral.  
The time derivative is symmetrized by integrating by parts and ignoring boundary terms. 
Thus, the full path integral between two states $\mid{\bf Q}'\rangle\mid\Psi^{0}\rangle$ and
$\mid{\bf Q}\rangle\mid\Psi^{f}\rangle$ is given by a total action which of the form:
\begin{equation}
S_{T}[{\bf R},\psi,\phi] = \int_{0}^{t}{\{L_{n}({\bf R},\dot{\bf R}) + \frac{\langle\phi({\bf R})\mid 
i\overleftrightarrow{\partial_{t}} - H_{e}\mid\psi({\bf R})\rangle}{\langle\phi({\bf R})\mid\psi({\bf R})
\rangle}\}dt}
\end{equation}
  Here the time derivative, $\overleftrightarrow{\partial}_{t}$, is the symmetrized derivative 
defined as:
\begin{equation}
{\langle\phi({\bf R})\mid i\overleftrightarrow{\partial_{t}}\mid\psi({\bf R})\rangle} =
\frac{i}{2}[{\langle\partial_{t}\phi({\bf R})\mid\psi({\bf R})\rangle} - 
\langle\phi({\bf R})\mid\partial_{t}\psi({\bf R})\rangle]
\end{equation}
 The purely nuclear contribution to the action integral is contained in the Lagrangian $L_{n}$ which 
is defined as follows:
\begin{equation}
  L_{n}({\bf R},\dot{\bf R}) = \frac{1}{2}M\dot{\bf R}^{2} - U({\bf R})
\end{equation}   
   Here the quantities $\bf R$ correspond to the set of nuclear coordinates, and $M$ is the 
mass of the nuclei. $U({\bf R})$ is the bare nuclear potential as defined earlier in Eq.(2). 
   The path integral in terms of this action can be written as
\begin{eqnarray}
\langle\Psi^{0};{\bf Q}'\mid G_{na}\mid\Psi^{f};{\bf Q}\rangle = \int{D(\psi^{*}({\bf R}(\tau),
\psi({\bf R}(\tau))D(\phi^{*}({\bf R}(\tau)),\phi({\bf R}(\tau))
\langle\Psi^{0}\mid\psi({\bf Q}')}\rangle\nonumber\\
\langle\phi({\bf Q})\mid\Psi^{f}\rangle\exp{\Big[iS_{T}[{\bf R}(\tau),\psi,\phi]\Big]}\nonumber\\
\end{eqnarray}

   Since, in general, off diagonal matrix elements of the time evolution operator in the coherent
state basis are being considered, the boundary conditions on the path integral are that the 
coherent states defined at the end points ${\bf Q}'$ and $\bf Q$ are allowed to vary without 
constraint. However, in the semiclassical limit, which corresponds to choosing only the subset 
of continuous paths in coherent state space, the coherent states at the boundaries are 
constrained to have unit overlaps with the initial and final electronic wavefunctions 
respectively. 
   
  As noted earlier, the necessity of choosing two sets of coherent states at each time slice 
in the path integral means that the paths in general are discontinuous. This is because, in 
the limit where the discrete time slicing in the path integral is made infinitesimally small,
the choice of different states at each such time slicing corresponds to a discontinuous jump 
in the space of coherent states. 
 
  However, the stationary phase limit of the path integral involves continuous paths in state
space. Hence, it is reasonable to reduce the unconstrained sum over all (discontinuous and 
continuous) paths to one where the coherent states are constrained to be similar to each other. 
A simple way of doing this is to assume the overlap, 
$\langle\phi({\bf R})\mid\psi({\bf R})\rangle$, to be nearly unity. This is a reasonable 
approximation also because contributions to the path integral when this overlap is small are 
negligible. This can be seen from the discrete version of the path integral, Eq.(13) and 
Eq.(16), from which it is clear that as the overlap goes to zero, the exponent in the path 
integral becomes highly oscillatory, and in addition is multiplied by the small overlap.  

  The action when the overlap is assumed to be unity can be considered to be a zeroth order 
approximation to the full action, when expanded in terms of the overlap. With this 
simplification, the path integral action becomes:
\begin{equation}
S_{T}[{\bf R},\psi,\phi] = \int_{0}^{t}{\big[L_{n}({\bf R},\dot{\bf R}) + \langle\phi({\bf R})\mid
i\overleftrightarrow{\partial}_{\tau} - H_{e}\mid\psi({\bf R})\rangle\big]d\tau}
\end{equation}
  This action has a simple interpretation. The time derivative part of the action corresponds 
to the nonabelian generalization of Berry's phase, while the other parts correspond to 
the dynamical evolution of the system. It is now possible to show that the generalized Berry's
phase in this action corresponds to nonadiabatic coupling terms between Born-Oppenheimer
energy levels. To show this, the Slater determinant states $\phi$ and $\psi$ are expanded 
in a local basis of Born Oppenheimer electronic eigenstates as below:
\begin{eqnarray}
\mid\psi({\bf R})\rangle = \sum_{\mu}{w_{\mu}[\psi,{\bf R}]\mid\Psi_{\mu};{\bf R}}\rangle\\
\mid\phi({\bf R})\rangle = \sum_{\mu}{w_{\mu}[\phi,{\bf R}]\mid\Psi_{\mu};{\bf R}}\rangle
\end{eqnarray}
The states $\mid\Psi_{\mu};{\bf R}\rangle$ are eigenstates of the electron Hamiltonian
\begin{equation}
[H_{e}({\bf r, R})]\mid\Psi_{\mu};{\bf R}\rangle = E_{\mu}({\bf R})\mid\Psi_{\mu};{\bf R}
\rangle
\end{equation}
  In this basis, the phase of the connection factor over the entire path, $\Gamma$ can be 
rewritten as  
\begin{equation}
\Gamma[C_{{\bf R}}] = \sum_{\mu\nu}{\int_{0}^{t}{d\tau\frac{1}{2}(w^{*}_{\mu}\langle\Psi_{\mu};{\bf R}\mid 
i\partial_{\tau}[w_{\nu}\mid\Psi_{\nu};{\bf R}\rangle] - i\partial_{\tau}[w^{*}_{\mu}
\langle\Psi_{\mu};{\bf R}\mid]w_{\nu}\mid\Psi_{\nu};{\bf R}\rangle)}}
\end{equation}
  This connection factor is the phase obtained by multiplying the connection factors $\Gamma_{k}$, 
in Eq.(20) over the entire path. A little algebra can be used to demonstrate that the connection factor 
can be further simplified into the following form:
\begin{equation}
\Gamma[C_{{\bf R}}] = \int_{0}^{t}{d\tau\sum_{\mu}{\frac{1}{2}(w^{*}_{\mu}[\phi,{\bf R}]
                 i\dot{w}_{\mu}[\psi,{\bf R}] +h.c)}} - \int_{C_{{\bf R}}}{d{\bf R}\cdot
\sum_{\mu\nu}{{\bf A}_{\mu\nu}({\bf R})w^{*}_{\mu}[\phi,{\bf R}]w_{\nu}[\psi,{\bf R}]}}
\end{equation} 
  Here, the coupling term ${\bf A}_{\mu\nu}$ is the familiar nonadiabatic coupling vector 
in the Born Oppenheimer basis:
\begin{equation}
{\bf A}_{\mu\nu} = -i\langle\Psi_{\mu}\mid\nabla_{\bf R}\Psi_{\nu}\rangle
\end{equation}

  The second term that contributes to the connection factor has the form of an integral 
over the gauge potential ${\bf A}_{\mu\nu}$. In contrast to the case of a simple Berry 
phase, this "gauge" potential is an infinite dimensional matrix and corresponds to a 
nonabelian geometric phase. Thus the nonadiabatic coupling between Born Oppenheimer surfaces 
has a purely geometric interpretation as arising due to the parallel transport of a local basis 
of electronic wavefunctions along a nuclear path. It is therefore a natural generalization 
of the Berry phase for parameter dependent quantum systems$\cite{berry}$. This 
generalized geometric phase also depends on the off-diagonal elements of the electronic 
density matrix, and consequently couples the slow, nuclear and fast, electronic degrees of 
freedom. In the limit where an infinite number of adiabatic states corresponding to the fast 
modes can be accessed, this phase represents a dissipative coupling between the system of 
slow nuclear modes and the bath of fast electronic modes$\cite{krishna}$.
  
   It should be noted that generalizations of Berry's phase to the case of nondegenerate 
electronic levels have previously been obtained by Mead$\cite{mead1}$ and by Zygelman
$\cite{zygelman1,zygelman2}$. A related path integral description of the nonadiabatic 
dynamics has also previously been derived by Moody, Shapere and Wilczek$\cite{shapere1}$. 
In their work, a path integral was obtained by introducing Born-Oppenheimer 
electronic eigenstates at each nuclear position, in the expression for $G_{na}[\bf R]$ given 
by Eq.(12). The electronic matrix elements of $G_{na}[\bf R]$ between initial and 
final Born Oppenheimer eigenstates $\mid\Psi_{m};{\bf Q}'\rangle$ and $\mid\Psi_{n};{\bf Q}\rangle$ 
are given by
\begin{eqnarray}
\langle\Psi_{m};{\bf Q}'\mid G_{na}[{\bf R}(t)]\mid\Psi_{n};{\bf Q}\rangle =
\sum_{m_{1},m_{2},..}{\exp[-i\epsilon E_{m}({\bf Q}')]\langle\Psi_{m};{\bf Q}'\mid
\Psi_{m_{1}};{\bf R}_{1}\rangle\exp[-i\epsilon E_{m_{1}}({\bf R}_{1})]}\nonumber\\
\langle\Psi_{m_{1}};{\bf R}_{1}\mid\Psi_{m_{2}};{\bf R}_{1}\rangle ....\langle
\Psi_{m_{N}};{\bf R}_{N}\mid\Psi_{n};{\bf Q}\rangle\nonumber\\ 
\end{eqnarray}
  This expression contains connection factors of the form:
\begin{equation}
\Gamma^{d}_{k} = \langle\Psi_{m_{k}};{\bf R}_{k}\mid\Psi_{m_{k+1}};{\bf R}_{k+1}\rangle   
\end{equation}
   It is easy to see that this connection factor can be manipulated as with the 
coherent state case to give 
\begin{equation}
\Gamma^{d}_{k} = [\delta_{mn} - \delta{\bf R}_{k}\cdot
\langle\Psi_{m_{k}};{\bf R}_{k}\mid\nabla_{{\bf R}_{k}}\Psi_{m_{k+1}};{\bf R}_{k}\rangle]
\end{equation}
  with $\delta{\bf R}_{k} = {\bf R}_{k+1} - {\bf R}_{k}$. 
  This expression can be rewritten in terms of the nonadiabatic coupling to give
\begin{equation}
\Gamma^{d}_{k} = \{\exp[-i\delta{\bf R}_{k}\cdot{\bf A}_{m_{k},m_{k+1}}({\bf R}_{k})]\}_
{m_{k},m_{k+1}}
\end{equation}
  This connection factor is similar to the connection factor, Eq.(20) derived in the 
coherent state electronic basis. It is however harder to handle for the purpose of 
approximations and computations, due to the discrete nature of the summation over 
Born-Oppenheimer eigenstates. Although the approach presented here is physically 
equivalent to that of Moody, Shapere and Wilczek, the use of electronic coherent states 
enables semiclassical approximations to the nonadiabatic quantum dynamics. This is 
possible because the restricted sum over only Born Oppenheimer eigenstates in Eq.(33) is 
replaced with an integral over continuous paths in the coherent state space. This provides 
a practical advantage, in that computationally tractable semiclassical approximations to the 
full path integral become possible. It is shown later in this work that stationary phase 
approximations to the nonadiabatic path integral so obtained, are closely related to the 
Ehrenfest formulation of mixed quantum-classical dynamics. 

  The path integral representation derived here is an exact reformulation of the quantum 
dynamics of the system. However, due to the continuum nature of the coherent state basis,
it is difficult to numerically evaluate the path integral propagator and obtain the full 
quantum dynamics.

\section*{III. Analysis}
  The path integral derived in the previous section can be used to develop a semiclassical
formulation of the nonadiabatic dynamics. In this section, stationary phase approximations 
to the path integral are derived and a semiclassical quantization rule is described for the 
nonadiabatic dynamics. The stationary phase equations so obtained form a natural quasiclassical
limit for the full quantum dynamics. For the coherent state path integral, this approximation 
is shown to yield the Ehrenfest equations of motion with an additional correction term. 
This correction term corresponds to the geometric contribution to the equations of motion, 
and is non-zero in general when the nuclear subspace is three dimensional or higher.

  For the sake of completeness, the stationary phase approximation to the coherent state 
path integral is compared to a generalized stationary phase type approximation to the 
nonadiabatic path integral when written out in the basis of electronic Born-Oppenheimer
eigenstates. It is shown that a generalization of the stationary phase approximation in 
this case yields a quasiclassical theory which is the target of the fewest switches
trajectory surface hopping method. A natural semiclassical extension of this approximation
yields a generalization of the fewest switches method which can account for hops in classically
forbidden regions. Finally, the derivation of general mixed quantum classical methods that 
combine the Ehrenfest and surface hopping methods is discussed. Thus, this section clarifies and 
lays down the neccesary theoretical foundations for the systematic improvement of mixed quantum 
classical methods.  
 
\subsection*{A. Quasiclassical equations of motion}
  A stationary phase approximation to the path integral can be made to yield quasiclassical 
equations of motion for the coupled dynamics of the nuclear and electronic degrees of 
freedom. Such approximations are derived here to obtain quasiclassical equations of motion
for the system. It is shown that the equations so obtained are a generalization of Ehrenfest's
equations of motion.  
  
  A density matrix for the electronic subsystem can be defined as 
\begin{equation}
\sigma_{\alpha\beta}[\psi,\phi;{\bf R}] = w^{*}_{\beta}[\phi,{\bf R}]w_{\alpha}[\psi,{\bf R}]
\end{equation}
  where the quantities $w_{\alpha},w_{\beta}$ are as defined in Eq.(27) and Eq.(28).
  To express the nonadiabatic action in terms of this density matrix, the explicit form of the 
action from Eqs.(25) and (30) is written to be 
\begin{equation}
S_{T}[{\bf R},\sigma] = \int_{0}^{t}{d\tau\Big[L_{n}({\bf R},\dot{\bf R}) + \frac{i}{2}
[w_{\mu}^{*}\dot{w}_{\nu} - \dot{w}_{\mu}^{*}w_{\nu}] - 
\dot{\bf R}\cdot\sum_{\mu\nu}{{\bf A}_{\mu\nu}w_{\nu}[\phi,t]w_{\mu}^{*}[\psi,t]}\Big]}
\end{equation}
  With this form of the action, the stationary phase approximation corresponds to the 
condition
\begin{equation}
\delta S_{T}[{\bf R},\sigma] = 0 .
\end{equation}
  A detailed derivation of the stationary phase equations is given in Appendix A. at 
the end of this work. On performing the appropriate manipulations, the stationary phase 
condition for the electronic density matrix, $\sigma$, can be shown to be the following 
equation of motion:
\begin{equation}
i\dot{\sigma}_{\alpha\beta} = (E_{\alpha}({\bf R})-E_{\beta}({\bf R}))\sigma_{\alpha\beta}
+ \dot{\bf R}\cdot\sum_{\gamma}{[{\bf A}_{\alpha\gamma}\sigma_{\gamma\beta} - 
\sigma_{\alpha\gamma}{\bf A}_{\gamma\beta}]}
\end{equation} 
   Here, the energies $E_{\alpha}({\bf R}),E_{\beta}({\bf R})$ are the eigenvalues of the Born
Oppenheimer eigenstates as defined in Eq.(29). The corresponding equation of motion for the nuclear 
coordinates, $\bf R$ with mass $M$ and the bare nuclear potential $U({\bf R})$ can be written to be 
\begin{equation}
M\ddot{\bf R} = -\nabla_{\bf R}U({\bf R}) -\sum_{\mu}{\nabla_{\bf R}E_{\mu}({\bf R})
\sigma_{\mu\mu}} + \sum_{\mu\nu}{{\bf A}_{\nu\mu}\dot{\sigma}_{\mu\nu}} -\sum_{\mu\nu}
{\sigma_{\mu\nu}\dot{\bf R}\times(\nabla_{\bf R}\times{\bf A}_{\nu\mu})} 
\end{equation}
 The stationary phase equation for the nuclear coordinates, $\bf R$, contains a term that is 
proportional to the time derivative of the off diagonal electron density matrix 
$\sigma_{\mu\nu}$. In addition to the other terms which are proportional to the off-diagonal
density matrix, this is an explicit contribution to the force acting on the nuclei due to 
the time variation of the electronic coherence. Hence, this accounts for changes in the 
effective nuclear force due to electronic dephasing.

  Eq.(41) can be further simplified and written explicitly in terms of the elements of the 
electronic density matrix, $\sigma$ and the nonadiabatic couplings $\bf A$ as, 
\begin{eqnarray}
M\ddot{\bf R} = -\nabla_{\bf R}V - \sum_{\mu}{\nabla_{\bf R}E_{\mu}\sigma_{\mu\mu}} + 
i\sum_{\mu\nu}{E_{\mu\nu}{\bf A}_{\mu\nu}\sigma_{\nu\mu}} - \sum_{\mu\nu}{\sigma_{\mu\nu}
\dot{\bf R}\times(\nabla_{\bf R}\times{\bf A}_{\nu\mu})} 
+ \nonumber\\
i\sum_{\mu\gamma\nu}{\sigma_{\nu\mu}\dot{\bf R}\times({\bf A}_{\mu\gamma}\times{\bf A}_{\gamma\nu})}
\end{eqnarray}
If a nonabelian "vector potential" is defined as an infinite dimensional matrix with elements
\begin{equation}
  [{\bf A}({\bf R})]_{\mu\nu} = {\bf A}_{\mu\nu}
\end{equation}
then the corresponding gauge invariant, nonabelian "magnetic field" is given by
\begin{equation}
  {\bf B}_{\mu\nu} = \nabla_{\bf R}\times{\bf A}_{\mu\nu} - i({\bf A}\times{\bf A})_{\mu\nu}
\end{equation}
From these definitions and Eq.(42), the stationary phase equations for the nuclear 
coordinates can be written in the form
\begin{equation}
M\ddot{\bf R} = -\nabla_{\bf R}V -\sum_{\mu}{\nabla_{\bf R}E_{\mu}\sigma_{\mu\mu}} +
i\sum_{\mu,\nu}{E_{\mu\nu}({\bf R}){\bf A}_{\mu\nu}\sigma_{\nu\mu}} - \dot{\bf R}\times\sum_{\mu,\nu}
{{\bf B}_{\nu\mu}\sigma_{\mu\nu}}
\end{equation}
   Eq.(40) corresponds to the familiar Liouville equation for the electronic density matrix,
while Eq.(45) is different because it contains corrections due to the nonadiabatic coupling.
The first three force terms correspond to the effective force in the Ehrenfest formulation
of mixed quantum classical dynamics, while the last term is a geometric correction to the 
Ehrenfest equations. This term therefore constitutes a new addition to the equations of motion 
and will need to be retained for consistency with the full quantum dynamics. When this term is 
ignored, the equations of motion so obtained correspond to the Ehrenfest equations of motion.

  The equation, Eq.(45), for the nuclear coordinates satisfies the invariance under
general unitary transformations of the electronic basis. To see this, note that the nonadiabatic
coupling matrix ${\bf A}({\bf R})$ is defined through the nonadiabatic coupling matrix
elements as
\begin{equation}
 [{\bf A}({\bf R})]_{\mu\nu} = {\bf A}_{\mu\nu}({\bf R}) = 
-i\langle\Psi_{\mu}\mid\nabla_{\bf R}\Psi_{\nu}\rangle
\end{equation}
  To write this in a more transparent form, infinite dimensional column vectors $\bf\Psi$ can be 
constructed with elements given by the electronic basis wavefunctions $\Psi_{\mu}$.
  The matrix can then be rewritten as a product of column vectors
\begin{equation}
{\bf A}({\bf R}) = {\bf\Psi}^{\dagger}(-i\nabla_{\bf R}){\bf\Psi}
\end{equation}
  It is straightforward to show that under a unitary transformation, $\Pi({\bf R})$  of the 
electronic wavefunctions,
\begin{equation}
{\bf\Psi}^{'} = \Pi({\bf R}){\bf\Psi}
\end{equation}
  the nonadiabatic coupling matrix transforms as 
\begin{equation}
  {\bf A}({\bf R}) \rightarrow \Pi^{\dagger}{\bf A}\Pi - i\Pi^{\dagger}\nabla_{\bf R}\Pi
\end{equation}
  The "magnetic field", ${\bf B}$ with matrix elements ${\bf B}_{\mu\nu}$ correspondingly
transforms as 
\begin{equation}
{\bf B} \rightarrow \Pi^{\dagger}{\bf B}\Pi
\end{equation}
  The density matrix elements $\sigma_{\nu\mu}$ also transform similarly under the unitary 
transformation, and hence the "magnetic field" contributions to the equations of motion are
invariant under such transformations. This invariance is a general version of the the usual 
invariance under gauge transformations, of vector potentials and magnetic fields in classical 
electromagnetism. It should be noted that the unitary transformations under which the electronic
wavefunction basis transforms are infinite dimensional matrices, and hence the group,
$U(\infty)$ that these unitary transformations belong to is the group of symmetries for 
the nonadiabatic corrections. On the other hand, the terms in the equations of motion
which are gradients of the electronic energy eigenstates break this symmetry, since
the energy eigenvalues are, in general, not completely degenerate. When degeneracies
exist for a subset of the electronic energy levels, the contribution to the nuclear
force from the degenerate states is symmetric under unitary transformations corresponding
to the degenerate subspace. The group corresponding to this symmetry is $U(n)$, where $n$
is the degree of the degeneracy. When this symmetry group exists, the time evolution of 
the degenerate electronic subspace picks up a Berry phase$\cite{zee}$ which contributes to the 
nuclear force in Eq.(45) in a fashion identical to that of the nonadiabatic coupling between
nondegenerate states. Thus, the equation of motion for the nuclei, Eq.(45), is valid 
irrespective of whether the states in the electronic subspace are degenerate or not. For 
example, in the case of a conical intersection between two potential energy surfaces, the 
wavefunction becomes multivalued. To resolve this multivaluedness, a geometric phase due 
to a vector potential can be added to the wavefunctions. The geometric forces in Eq.(45) 
are the classical analog of this procedure.        

   It should be noted that the geometrical, "magnetic field" contributions in Eq.(45) are 
zero for one dimensional nuclear potential energy surfaces. This can be intuitively understood 
through the discussion in Sec.II.A. This magnetic contribution is due to the "rotation" of 
the local electronic Hilbert space along the nuclear path. For one dimensional nuclear potential 
energy surfaces, this "rotation" does not have any torsional components, and is simply a pure 
parallel transport of the basis. Thus, for single nuclear dimensions, these forces are zero and 
do not contribute to the dynamics. However, for potential energy surfaces parametrized by higher 
dimensional nuclear coordinates, these forces are non-zero and of importance to the nonadiabatic 
dynamics. These couplings therefore cause an effective magnetic force contribution to be added to 
the usual Ehrenfest force, in the mixed quantum classical dynamics. The magnetic forces are the 
quasiclassical analog of the additional Berry-Mead phase which the electronic wavefunctions are 
multiplied by, to maintain their single valuedness.

\subsection*{B. The Discrete Path Integral}
   A complementary description of the full nonadiabatic quantum propagation of the nuclei 
can be obtained by writing the nonadiabatic propagation operator $G_{na}[{\bf R}]$ from 
Eq.(12) in a basis of electronic Born Oppenheimer eigenstates. As mentioned previously 
such a path integral expression for the propagator has been written down by Moody, Shapere
and Wilczek. Explicitly, if the wavefunction for the combined system is given by
\begin{equation}
\langle{\bf Q}\mid\Psi,t\rangle = \sum_{m}{\xi_{m}({\bf Q},t)\mid\Psi_{m};{\bf Q}\rangle}
\end{equation}
then the time evolution of the nuclear states $\xi_{m}$ can be written as
\begin{equation}
\xi_{m}({\bf Q}',t) = \sum_{n}{\int{K_{mn}({\bf Q}',t\mid{\bf Q},0)\xi_{n}({\bf Q},0)d{\bf Q}}}
\end{equation}
The propagator $K_{mn}$ is the matrix element of the time evolution operator when evaluated 
between an initial electronic Born-Oppenheimer state, $\Psi_{n}$, defined for an initial 
position $\bf Q$ and a final electronic Born-Oppenheimer state, $\Psi_{m}$, defined for a 
final nuclear position $\bf Q'$. This matrix element can be evaluated using the Trotter product 
formula to slice the time $t$ into infinitesimal time intervals, and introducing a basis of 
electronic Born Oppenheimer states at each time slice. The resulting propagator is the same as  
the propagator defined in Eq.(33).
  
Explicitly, this propagator is the matrix element 
\begin{equation}
 K_{mn}({\bf Q}', t\mid{\bf Q}, 0) = \langle\Psi_{m};{\bf Q}'\mid e^{-iHt}\mid\Psi_{n};{\bf Q}\rangle
\end{equation}
 If the Born-Oppenheimer limit is assumed to be valid, then the propagator is diagonal in the indices 
$m,n$. 
\begin{equation}
K^{b.o}_{mn}({\bf Q}', t\mid{\bf Q}, 0) = \delta_{mn}K^{b.o}_{mm}({\bf Q}', t\mid{\bf Q}, 0)
\end{equation}
Here, $K^{b.o}_{mm}$ is the nuclear propagator in a potential given by $V_{m}({\bf Q}) = U({\bf Q}) + 
E_{m}({\bf Q})$
\begin{equation}
K_{mm}({\bf Q}', t\mid{\bf Q}, 0) \approx \langle{\bf Q}'\mid e^{-i(T_{n} + V_{m}({\bf R}))t}\mid{\bf Q}\rangle
\end{equation} 
The nonadiabatic nuclear propagation can be thought of as a sequence of propagation along 
Born-Oppenheimer paths with the propagator defined above in Eq.(54), and "hops" occurring in 
between to switch between Born-Oppenheimer paths. This is illustrated in Fig.2. Detailed 
considerations of the structure of the paths that contribute to the transition 
amplitude $K_{mn}$ thus point to the origins of surface hopping type approximations. 

To derive explicitly quasiclassical techniques based on the discrete path integral, consider the 
effective propagation operator defined in Eq.(12), which corresponds to matrix elements of the full 
propagation operator between initial and final nuclear position eigenstates. To recapitulate,
the matrix elements are given as an integral of a path ordered exponential $G_{na}[{\bf R}(t)]$. For a
given nuclear path ${\bf R}(t)$ the matrix elements of $G_{na}$ between an initial Born-Oppenheimer
electronic state $\Psi_{n}$ and a final state $\Psi_{m}$ are given by
\begin{equation}
\langle\Psi_{m};{\bf Q}'\mid G_{na}[{\bf R}(t)]\mid{\bf Q};\Psi_{n}\rangle = 
\langle{\bf Q}';\Psi_{m}\mid e^{-iH({\bf Q}')\epsilon}e^{-iH({\bf R}_{1})\epsilon}...e^{-iH({\bf Q})
\epsilon}\mid{\bf Q};\Psi_{n}\rangle 
\end{equation}
From this relationship, the quantum nonadiabatic propagator between an initial Born oppenheimer 
electronic wavefunction $\mid\Psi_{n};{\bf Q}\rangle$ defined at an initial nuclear position 
$\bf Q$ and a final Born Oppenheimer electronic wavefunction, $\mid\Psi_{m};{\bf Q}'\rangle$ defined 
for a position $\bf Q'$ can be written as 
\begin{equation} 
   K_{mn}({\bf Q}',t\mid{\bf Q},0) = \int{D{\bf R}(\tau)\langle{\bf Q}';\Psi_{m}\mid 
G_{na}[{\bf R}(t)]\mid{\bf Q};\Psi_{n}\rangle} 
\end{equation}
  If a nonadiabatic transition takes place during the interval $(t',t'+\epsilon)$ for sufficiently 
small values of $\epsilon$, the propagator has the composition property
\begin{equation}
K_{mn}({\bf Q}',t\mid{\bf Q},0) = \int{d{\bf R}'d{\bf R}''\sum_{pq}{ K_{mp}({\bf Q}',t\mid{\bf R}',t')
\langle\Psi_{p};{\bf R}'\mid\Psi_{q};{\bf R}''\rangle K_{qn}({\bf R}'',t' + \epsilon\mid{\bf Q},0)}}
\end{equation}
 The overlap between electronic wavefunctions at ${\bf R}'$ and ${\bf R}''$ is the connection factor,
$\Gamma_{pq}({\bf R}',{\bf R}'')$, discussed in the previous section, and is responsible for the 
nonadiabatic coupling between levels. As before, it can be rewritten as:
\begin{equation}
 \langle\Psi_{p};{\bf R}'\mid\Psi_{q};{\bf R}''\rangle = \Gamma_{pq}({\bf R}',t\mid{\bf R}'',t') \approx 
(\delta_{pq} - i{\bf A}_{pq}\cdot({\bf R}' - {\bf R}''))
\end{equation}
 with the nonadiabatic coupling ${\bf A}_{mn}$ given by Eq.(32) as before. 
 This composition law for the exact nonadiabatic propagator can be used to derive a Born series 
expansion for the propagator in terms of the Born Oppenheimer propagator. The physics behind this 
is illustrated in Fig.2. 
\begin{eqnarray}
K_{mn}({\bf Q}',t\mid{\bf Q},0) = K^{b.o}_{mm}({\bf Q}',t\mid{\bf Q},0) + \int{d{\bf R}'K^{b.0}_{mm}
({\bf Q}',t\mid{\bf R}',t')\Gamma_{mn}({\bf R}',{\bf R}'')
K^{b.o}_{nn}({\bf R}'',t'+\epsilon\mid{\bf Q},0)}\nonumber\\
  + \textit{terms with two nonadiabatic transitions}  + .... \nonumber\\
\end{eqnarray}
  Here, the propagator is written as a sum over paths where there are successively higher number of 
nonadiabatic transitions. To obtain a classical nuclear propagation scheme, the nuclei can be 
propagated along the stationary phase solutions for those pieces of the nuclear path where the 
propagator is of the Born Oppenheimer type. In other words, the Born-Oppenheimer limit of the 
nonadiabatic propagator can be approximated with a semiclassical expansion around the stationary 
phase path as discussed below.  

  It is easy to see that along a Born Oppenheimer path,
\begin{equation}
i\frac{\partial\xi_{m}}{\partial t} = H_{m}\xi_{m}
\end{equation}
  Furthermore, if a single nonadiabatic transition occurs from $n\rightarrow m$ on a sufficiently
short timescale $\epsilon$, the amplitude $\xi_{m}$ changes by
\begin{equation}
\delta\xi_{m}({\bf R}',t) = -i\sum_{n}{\delta{\bf R}'\cdot{\bf A}_{mn}({\bf R}')\xi_{n}({\bf R}',t)}
\end{equation}
 Putting the two equations together, the total time derivative of the wavefunction $\xi$ is recovered
to be:
\begin{equation}
i\dot{\xi}_{n}({\bf R},t) = H_{n}\xi_{n} + \dot{\bf R}\cdot\sum_{p}{{\bf A}_{np}({\bf R})\xi_{p}({\bf R},t)}
\end{equation}
  It is important to note that this equation is valid under the assumption that nonadiabatic 
transitions occur on a very short timescale in comparison to the timescales that govern the 
rest of the dynamics. In general, there is a nonzero probability for nonadiabatic transitions 
occuring over longer timescales and accompanied by finite changes, $\delta{\bf R}'$ in the nuclear 
coordinates. This equation is derived assuming that the contribution from such transitions is 
small, and that dominant features of the nonadiabatic dynamics can be captured by considering 
transitions that occur on infinitesimal timescales. 
 
  To derive a semiclassical limit for the nuclear evolution, the nuclear wavefunctions 
$\xi_{n}({\bf R},t)$ are approximated as Gaussian wavefunctions centered around 
the instantaneous classical nuclear path on the Born Oppenheimer surface, according to 
the prescription of Heller$\cite{Heller}$. 
\begin{equation}
\xi_{n}({\bf R},t) \approx c_{n}(t)F_{n}[{\bf R}_{n}(t),{\bf P}_{n}(t)]
\end{equation}
  The functions $F_{n}(t)$ are instantaneous Gaussian wavepackets centered around the
classical Born-Oppenheimer trajectories given by
\begin{eqnarray}
\frac{d{\bf R}_{n}}{dt} = \frac{\partial H_{n}}{\partial{\bf P}}; {\bf P} = {\bf P}_{n}(t)\nonumber\\
\frac{d{\bf P}_{n}}{dt} = - \frac{\partial H_{n}}{\partial{\bf R}}; {\bf R} = {\bf R}_{n}(t)
\end{eqnarray}
 Furthermore, this Gaussian wavepacket approximately satisfies (when the Born Oppenheimer 
potential energy, $U({\bf R}) + E_{n}({\bf R})$, is expanded to quadratic order in ${\bf R}$) the 
time dependent Schrodinger equation:
\begin{equation}
i\dot{F}_{n} \approx H_{n}F_{n}
\end{equation}
  Substituting the ansatz, Eq.(64), into Eq.(63), one obtains for the time evolution 
of the coefficients, $c_{n}$
\begin{equation}
i\dot{c}_{n}F_{n} + ic_{n}\dot{F}_{n} = c_{n}H_{n}F_{n} + \sum_{p}{\dot{\bf R}\cdot{\bf A}_{np}F_{p}c_{p}}
\end{equation}
 Making use of the approximation in Eq.(66), the time evolution of $c_{n}$ becomes, in the 
semiclassical limit:
\begin{equation}
 i\dot{c}_{n}F_{n} = \sum_{p}{\dot{\bf R}\cdot{\bf A}_{np}F_{p}c_{p}}
\end{equation}
  Multiplying by $F^{*}_{n}$ and integrating over the coordinates ${\bf R}$ gives
\begin{equation}
   i\dot{c}_{n} = \sum_{p}{\langle F_{n}\mid\dot{\bf R}\cdot{\bf A}_{np}\mid F_{p}\rangle c_{p}}
\end{equation}
  This can be rewritten in terms of a density matrix $\sigma_{nm} = c_{n}c^{*}_{m}$ as 
\begin{equation}
   i\dot{\sigma}_{nm} = \sum_{p}{[\langle F_{n}\mid\dot{\bf R}\cdot{\bf A}_{np}\mid F_{p}\rangle\sigma_{pm} -
     \sigma_{np}\langle F_{p}\mid\dot{\bf R}\cdot{\bf A}_{pm}\mid F_{m}\rangle]}
\end{equation}
   Furthermore, if the Gaussians $F_{n}$ are assumed to be delta functions in the positions $R$, the 
above equations simplifies further into:
\begin{equation}
i\dot{\sigma}_{nm} \approx \sum_{p}{[\dot{\bf R}\cdot{\bf A}_{np}\sigma_{pm} - 
\sigma_{np}\dot{\bf R}\cdot{\bf A}_{pm}]}
\end{equation}
 where the variable $\bf R$ and its velocity, $\dot{\bf R}$ are classical 
quantities corresponding to the center of the Gaussian function $F_{n}$. 
Since the functions $F_{n}$ are Gaussians centered on the classical nuclear
Born-Oppenheimer trajectories, the classical variables $\bf R$ satisfy Newton's laws:
\begin{equation}
  M\ddot{\bf R} = - \nabla_{\bf R}[U({\bf R}) + E_{n}({\bf R})]
\end{equation}
when the trajectory being considered is that belonging to the $n^{th}$ Born-Oppenheimer surface.
Further, the rate of change of the diagonal elements, $\sigma_{nn}$ is given by
\begin{equation}
  i\dot{\sigma}_{nn} = \dot{\bf R}\cdot\sum_{p}{[{\bf A}_{np}\sigma_{pn} - \sigma_{np}{\bf A}_{pn}]}
\end{equation}
  It is clear from this equation and the fact that the nuclear propagation always occurs on a single 
Born-Oppenheimer surface, that these equations describe the surface hopping method$\cite{tully1,tully2}$
, for which the fewest switches algorithm is a Monte Carlo solution. Thus, the fewest switches surface 
hopping (FSSH) approach$\cite{tully2}$ corresponds to a semiclassical approximation to the exact quantum 
nonadiabatic dynamics, similar to the Ehrenfest method described in Sec III.A.

  It is also interesting to note that Eq.(70) can be used instead of Eq.(73) to generate the surface
hopping criterion. Eq.(70) provides for a nonzero hopping probability even in regions where
transitions are classically forbidden, unlike the usual FSSH method. A similar extension has 
also been proposed previously by Neria and Nitzan$\cite{Neria}$, on physical grounds through the construction
of a semiclassical golden rule. In addition to this, it is clear from this discussion that semiclassical 
propagation of the nuclear wavefunction, either using the Gaussian form, Eq.(65) and Eq.(66) proposed by 
Heller$\cite{Heller}$, or by the use of a different basis, lead to generalized schemes for nonadiabatic 
dynamics that go beyond the mixed quantum classical limit. The use of frozen gaussians within this framework 
correspond to a scheme for nonadiabatic dynamics that is similar to the multiple spawning method proposed by 
Martinez and co-workers$\cite{Martinez,Martinez1}$.  
 
  The approximation of hops occurring over infinitesimal timescales that is required to derive 
both Eq.(70) and Eq.(73) causes phase changes in the wavefunction due to the hops to be neglected. 
Furthermore, hops need not occur on infinitesimal timescales, and the quantum nuclear paths which 
contribute to the dynamics are in general not differentiable, rendering the nuclear velocity to be 
a poorly defined quantity. In circumstances where such paths contribute significantly to the dynamics
surface hopping type methods will be invalid. Additionally, in the case where two (or more) electronic 
potential energy surfaces become degenerate, the nonadiabatic coupling also picks up nonzero diagonal 
elements which could act on the dynamics along significant sections of the nuclear path. In such 
situations, an approximate means of treating the dynamics would be to add the diagonal terms in the 
nonadiabatic coupling to the nuclear Hamiltonian and perform the nuclear propagation on each Born 
Oppenheimer surface with an additional magnetic force due to the diagonal terms in the nonadiabatic 
coupling. The off diagonal nonadiabatic couplings can then still be treated as acting instantaneously
to cause hops. It is likely that such modifications could be of value in improving the accuracy of the 
FSSH method in such situations.

  Finally, it is useful to discuss the relative utility of the discrete basis path integral presented
here versus the coherent state approach that leads to the Ehrenfest equations in the previous section. 
The natural stationary phase limit of the coherent state path integral is the mean field Ehrenfest 
theory, while the discrete path integral has a corresponding stochastic limit. These two limits 
can be thought of as corresponding to two different types of statistics in the hopping events. 
The coherent state path integral is a better starting point for approximations when there is 
significant nonadiabatic coupling between multiple Born-Oppenheimer surfaces, i.e when hops can 
occur with high frequency. In such a situation, a mean field treatment is a good first approximation 
to the dynamics. However, the discrete basis approach is better suited to studying the dynamics 
in the presence of nonadiabatic dynamics between a few levels, and when Born Oppenheimer surfaces 
begin to diverge. In such a situation, the hopping events are relatively infrequent, and nonadiabatic 
behaviour is sensitive to the detailed structure of neighbouring Born-Oppenheimer surfaces. The 
discrete basis approach, for which the surface hopping algorithm is an approximation is a good 
starting point for approximation in this situation. 
     
\subsection*{C. Semiclassical quantization conditions}
  The nonadiabatic action Eq.(38) can be used to describe a semiclassical quantization 
condition when the system is in a bound state. This condition is derived in a fashion 
similar to that used to study the semiclassical quantization of coherent state path 
integrals$\cite{NegOr,Iida}$, and in quantum field theories$\cite{neveu1}$. 

  If an assumption is made that the system executes periodic orbits, then a quantization 
condition can be written. To derive this quantization condition, the trace of the propagator 
is written as 
\begin{equation}
  K_{na}(T) = Tr[\exp{(-iHT)}]
\end{equation}
  The trace is related to the density of states as below. 
\begin{equation}
  \rho(E) = Tr\Big[\frac{1}{E - H + i\eta}\Big] = i\int_{0}^{\infty}{e^{iET}K_{na}(T) dT}
\end{equation}
   If the trace corresponding to $\rho(E)$ is expanded in terms of the eigenstates 
of the entire (the combined nuclear and electronic) system, $\rho(E)$ can be written
in terms of the system energy eigenvalues $E_{N}$ as 
\begin{equation}
\rho(E) = \sum_{N}{\frac{1}{E - E_{N} + i\eta}}
\end{equation}
   It is clear from this that the poles of the function $\rho(E)$ correspond to the 
energy eigenvalues of the system. Hence, the poles of the Fourier transform of the 
propagator trace are related to the eigenstates of the combined system in this fashion. 

   The Fourier transform of the trace of the propagator can be written in terms of the
nonadiabatic coherent state path integral as follows:
\begin{equation}
  \rho(E) = i\int_{0}^{\infty}dT\int{d{\bf R}d\Psi\int{D[{\bf R}(t)]D[\psi(t)]
\exp{(iS_{T}[{\bf R},\psi,\phi] + ET)}}}
\end{equation}
   Here, the quantities $\psi,\phi$ are coherent state basis elements and the action 
$S_{T}$ is the same as defined in Eq.(26). To obtain an approximate quantization rule,
only diagonal coherent state matrix elements in Eq.(26) are considered.  
   Eq.(75) which relates the Fourier transform $\rho(E)$ to the energy spectrum of the 
system also implies that $\rho(E)$ is singular when the energy $E$ is equal to any 
of the system eigenenergies. This implies that the path integral of the propagator
in Eq.(77) is a maximum for such values of the energy $E$. The propagator also 
acquires its maximal values when its phase is stationary. Consequently, the density of 
states, $\rho(E)$, is dominated by contributions to the path integral from stationary
phase paths.  

   Furthermore, if it is assumed that periodic orbits exist for the combined dynamics of 
the system, then the density of states in Eq.(75) can be approximately written as 
\begin{equation}
  \rho[E] = \sum_{p.o}{C_{k}\exp{[iS[T_{k}(E)] + iET_{k}(E)]}}
\end{equation}
 where the time $T(E)$ is the time period for the classical traversal of the $k$th 
periodic orbit. Since, multiple traversals of a given periodic orbit also constitutes
a periodic orbit, this sum can be simplified into a sum over periodic orbits $\mu$ which 
cannot be reduced further into smaller sets of periodic orbits:
\begin{equation}  
\rho(E) = \sum_{\mu}{-iC_{\mu}\frac{\exp{[iS[T_{\mu}(E)] + iET_{\mu}(E)]}}{1 - 
\exp{[iS[T_{\mu}(E)] + iET_{\mu}(E)]}}}
\end{equation}
  This expression has poles given by 
\begin{equation}
 2\pi (n + \frac{1}{2}) = ET(E) + S[T,E]
\end{equation}
 This quantization condition can be explicitly written out to be 
\begin{equation}
  ET_{n}(E) + \int_{0}^{T_{n}}{dt \Big[L_{n}({\bf R,\dot{R}}) + \langle\psi\mid 
i\partial_{t} - H_{e}\mid\psi\rangle\Big]} = 2\pi(n + \frac{1}{2})
\end{equation}
  It should be noted that this quantization condition is valid under the assumption that
periodic orbits corresponding to the effective classical system defined by the action $S$
exist. For nuclear motion coupled to a large number of electronic levels, finding the 
periodic orbits of the system is a difficult task. However, for small values of the nonadiabatic 
coupling, this quantization rule provides an approximate means of estimating the total energy 
spectrum of the system, given an initial set of Born-Oppenheimer surfaces. Furthermore, if the 
nonadiabatic transitions couple only a finite set of electronic Born-Oppenheimer surfaces, this 
rule may have validity because periodic orbits for the combined nuclear and electronic density 
matrix variables could exist. 

  This rule can be written in terms of the total connection factors $\Gamma[C_{\bf R}]$
for the closed path $C_{\bf R}$ as 
\begin{equation}
   \oint{{\bf P}\cdot d{\bf R}} + \Gamma[C_{\bf R}] = 2\pi(n + \frac{1}{2})
\end{equation}

  A special case of the Born Oppenheimer problem, for which the semiclassical quantization
rule is useful, is that of a single vibrational coordinate, $\bf p,q$, coupled to a set of 
electronic levels. In this situation, the quantization rule reads
\begin{equation}
\oint{{\bf p}\cdot d{\bf q}} = 2\pi(n + \frac{1}{2}) - \oint{\sum_{k,l}{{\bf A}_{kl}({\bf q})\sigma_{lk}({\bf q})}\cdot d{\bf q}}
\end{equation}
 In the absence of the nonadiabatic couplings, ${\bf A}_{nm}$, the quantization rule gives the 
exact vibrational energy levels (assuming that the vibrations are purely harmonic). The 
contribution due to nonzero nonadiabatic coupling  can be interpreted as a shifting of the 
energy levels of the oscillator to account for energy exchange with the electronic bath. If an 
orbit is executed, nonadiabatic transitions between vibrational states can occur, which leads to 
energy transfer between the oscillator and the electronic system. 

\section*{IV. Conclusions}
     This work has presented an exact first principles path integral formulation of 
quantum nonadiabatic dynamics. The path integral approach developed here has the 
advantage that it enables computationally tractable stationary phase approximations 
to the full path integral. Previous path integral based approaches to this problem 
have stationary phase approximations that lead to nonlocal theories of mixed quantum 
classical dynamics$\cite{pech1,pech2}$, or require considering a constrained set of 
paths for the purposes of approximation$\cite{shapere1,shapere2}$. The method presented 
here therefore represents an advance over previous approaches in this regard. 

    A second consequence of this work is that corrections to the quasiclassical nuclear 
dynamics have been obtained. These corrections reflect underlying geometric features of 
the electronic Hilbert space. The corrections so obtained have the advantage that 
they can be calculated through straightforward extensions of current mixed 
quantum-classical simulation methodologies. Applications of this approach to 
study the dynamics of real systems in the presence of conical intersections and 
other degeneracies in the electronic energy levels are currently being considered.
Finally, semiclassical quantization rules have been proposed for nonadiabatic
dynamics in the quasiclassical limit. These rules are potentially of utility in 
determining approximate energy spectra when the Born-Oppenheimer approximation is not 
valid. 

   A third consequence of this work has been to provide a link between exact quantum 
theories of nonadiabatic dynamics and semiempirical surface hopping methods. This 
work establishes a rigorous theoretical basis for the surface hopping method. It has 
been demonstrated that the fewest switches surface hopping approach can be regarded
as a generalized stationary phase approximation to the exact quantum dynamics, in 
complement to the Ehrenfest formulation of mixed quantum classical dynamics. Natural
extensions of the surface hopping method to include classically forbidden nonadiabatic 
transitions, as well as the inclusion of nuclear quantum effects through the construction
of rigorous multiple spawning type algorithms have been demonstrated. This work also 
allows for the construction of semiclassical golden rules to 
describe nonadiabatic transitions. 
 
   The present work also provides a clear perspective on issues concerning equilibrium 
between the quasiclassical nuclear and quantum, electronic subsystems. It is clear from 
the theory presented here that traditional Ehrenfest methods do not satisfy equilibrium 
conditions if only single mean field trajectories are considered. Each such trajectory
corresponds to a stationary phase solution for the nonadiabatic propagator at a given set
of initial conditions. However, thermal properties of the system enter through sampling of 
initial conditions. Hence, if appropriately sampled ensembles of mean field trajectories 
are chosen, a mixed quantum-classical equilibrium can be established. Correct sampling of 
mean field trajectories will require an initial sampling of the reduced density matrix 
for the nuclei and should also correctly account for initial phase correlations according
to Eq.(26). It is possible that observed violations of mixed quantum classical equilibrium 
by the Ehrenfest method are due to an incorrect sampling of either of these contributions.
Since the surface hopping approximation is obtained by looking at the propagation of the system 
in each of the states occupied by it, an ensemble averaging of surface hopping trajectories will
sample the initial density matrix of the system, and hence this approach and any systematic 
generalizations based on it can be expected to reproduce the correct mixed quantum classical 
equilibrium. This has been demonstrated through a separate set of arguments$\cite{priya}$.
 
   This work provides a different conceptual perspective on the nature of nonadiabatic 
dynamics. Nonadiabatic couplings can be regarded as geometrical objects which are 
generalizations of Berry's geometric phase. Thus, they can be regarded on the same 
footing as the geometrical phases that arise due to conical intersections$\cite{vleck,longuet,truhlar}$. 
This similarity between nonadiabatic couplings and geometric phases that arise in degenerate 
systems has also been noted by previous authors$\cite{shapere1,Iida}$. In principle, the 
methods described in this work do not differentiate between systems which have degeneracies 
in energy and those which are nondegenerate. Therefore, the nonadiabatic path integral 
formulation presented here is of considerable generality and may be of practical utility 
in computational studies of nonadiabatic dynamics. 
 
   A potentially useful approach to study exact quantum nonadiabatic dynamics is to consider 
quantum Monte Carlo schemes to evaluate the coherent state path integral. Direct applications 
of quantum Monte Carlo schemes are likely to face difficulties due to the sign problem. 
However, simplifications could be made by using an effective single electron ansatz to 
replace the matrix elements in the action Eq.(26). Such an ansatz could be rigorously 
constructed by using the Runge-Gross theorem$\cite{Runge}$ which relates the exact one 
electron time dependent density to the expectation value of the electronic Hamiltonian 
in Eq.(26). Extensions of the time dependent density functional approach to one electron 
density matrices can then be used to accurately evaluate the path integral due to the action 
in Eq.(26). This has the added advantage of naturally incorporating time dependent 
density functional theory with the nonadiabatic propagation of a quantum system. This 
approach will be explored in later work. 

   Although the exact coherent state path integral presented here is a challenge
for numerical computation, it is possible to develop numerically more tractable 
semiclassical propagators that describe the nonadiabatic dynamics as an approximation 
to the coherent state path integral. One such approach would involve a stationary 
phase evaluation of the electronic dynamics (the Ehrenfest equations for the electron 
density matrix), while propagating the nuclear motion using a semiclassical initial 
value representation. More promising applications of this work are in developing better 
approximate methods to describe quantum nonadiabatic dynamics. One potential application 
is to develop  methods which systematically combine Ehrenfest and surface hopping type 
algorithms. To enable this the propagation of the nuclear dynamics can be divided into 
sections wherein different basis sets are used depending on physical considerations 
regarding the nature of nonadiabatic interactions along the nuclear paths. Another 
natural extension is in refining the surface hopping method to include classically 
forbidden nonadiabatic transitions by approximating the nuclear wavefunction with a 
gaussian wavepacket. Alternatively, more general semiclassical propagation schemes 
can be used to propagate nuclear dynamics in the discrete representation of the path 
integral. This would facilitate the inclusion of nuclear quantum effects as well as 
more accurate treatment of dynamics in the presence of strong nonadiabatic coupling 
between Born-Oppenheimer surfaces.

    The use of coherent states, and the derivation of the path integral presented here is
analogous to path integral treatments of quantum spin systems. This analogy with quantum spin
systems also forms the basis for classical mapping techniques$\cite{miller1,miller2,miller3,
stock,thoss}$ that have been developed to study mixed quantum classical dynamics. Finally, 
the path integral methods presented here are apparently amenable to perturbative approximations 
around its stationary phase value. However difficulties remain due to the nature of the 
nonadiabatic corrections to Born Oppenheimer dynamics. The nonadiabatic coupling between Born 
Oppenheimer levels consists of matrix elements of the nuclear momentum operator. The nuclear 
momentum operator is unbounded in the Hilbert space, and further, the nuclear paths that 
contribute to the path integral for the system are in general not differentiable. Hence, a 
perturbation expansion in the nonadiabatic coupling element is not well defined in the path 
integral framework$\cite{shapere1}$. In the case of Ehrenfest dynamics, this issue also manifests
through the assumptions made in constructing the coherent state path integral. The assumption made 
in coherent state path integrals is usually that the paths chosen in the coherent state space are
continuous. This is not valid in general for the problem of nonadiabatic dynamics. Similarly, 
extensions of surface hopping type methods will need to account for non-differentiable nuclear 
paths where the nuclear velocity is not defined. A means of overcoming this problem for the 
coherent state path integral is to regularize the path integral by considering higher order time 
derivatives in the action, in analogy to the procedure followed for spin systems$\cite{morandi,stone}$. 
A perturbative treatment of nonadiabatic dynamics that takes these issues into account is being 
worked on and will be presented at a later date.  

\section*{V. Acknowledgements}
     The major part of this work was done when the author was at Yale University. 
Financial support from a Yale University Graduate teaching fellowship is gratefully 
acknowledged. The author thanks Professor J.C Tully for his support and encouragement, and for 
discussions that motivated this work. The author acknowledges Dr.A.V Madhav and Professor J. 
Simons for critical comments and suggestions regarding the manuscript. The author is grateful 
to an anonymous referee for remarks and criticism which greatly improved and clarified 
the content of this paper. 

\renewcommand{\theequation}{A\arabic{equation}}
\setcounter{equation}{0}
\section*{Appendix A. Coherent state path integral: Stationary phase equations.}
   The derivation of the stationary phase equations in Sec.III A. is described here.
The effective action in the coherent state representation is given by
\begin{equation}
S_{T} = \int_{0}^{t}{L_{T}d\tau}
\end{equation}
 with a Lagrangian $L_{T}$ given by
\begin{equation}
L_{T} = \sum_{I=1}^{M}{\frac{M\dot{\bf R}_{I}^{2}}{2}} - U({\bf R}) - 
\sum_{m}{E_{m}({\bf R})w^{*}_{m}w_{m} + \frac{i}{2}[w^{*}_{m}\dot{w}_{m} -\dot{w}^{*}_{m}w_{m}]}
  - \dot{\bf R}\cdot\sum_{mn}{{\bf A}_{mn}w^{*}_{m}w_{n}}
\end{equation}
  For the stationary phase path, the action is stationary when varied with respect to the 
variables $\bf R$ and $w_{m},w^{*}_{m}$ which determine the path. In other words 
the stationary phase condition:
\begin{equation}
\delta S_{T} = 0
\end{equation}
 implies the equations:
\begin{eqnarray}
\frac{d}{dt}\frac{\partial L_{T}}{\partial\dot{\bf R}} - 
\frac{\partial L_{T}}{\partial{\bf R}} = 0 \\
\frac{\delta L_{T}}{\delta w_{m}} = 0 ;
\frac{\delta L_{T}}{\delta w^{*}_{m}} = 0
\end{eqnarray}
  The first of the three equations above is the usual Lagrangian equation of motion 
for Newtonian mechanics which governs the classical dynamics of the variable ${\bf R}(t)$,
while the next two equations render the action stationary with respect to variations of 
the electronic density matrix. The equations for the variables $w_{n},w^{*}_{n}$ are obtained
by using the appropriate unsymmetrized form of their kinetic energy term (the fourth term in
the right hand side of Eq.(A2)).
  
  On applying these equations of motion to the Lagrangian, one obtains
\begin{eqnarray}
   \frac{d}{dt}\{M\dot{\bf R} - \sum_{m,n}{{\bf A}_{mn}\sigma_{nm}}\} = -\nabla_{\bf R}U - 
\sum_{m}{\sigma_{mm}\nabla_{\bf R}E_{m}({\bf R})} - \sum_{m,n}{\nabla_{\bf R}(\dot{\bf R}\cdot{\bf A}_{mn})\sigma_{nm}}\\
-i\dot{w}^{*}_{n} = \dot{\bf R}\cdot\sum_{m}{w^{*}_{m}{\bf A}_{mn}} + E_{n}({\bf R})w^{*}_{n}\\
i\dot{w}_{n} = \dot{\bf R}\cdot\sum_{m}{{\bf A}_{nm}w_{m}} + E_{n}({\bf R})w_{n}
\end{eqnarray}
The second and third equations, Eq.(A7) and Eq.(A8), can be combined to give the Liouville Von-Neumann
equations for the electronic density matrix, $\sigma_{nm}=w_{n}w^{*}_{m}$:
\begin{equation}
i\dot{\sigma}_{nm} = (E_{n} - E_{m})\sigma_{nm} + \dot{\bf R}\cdot\sum_{k}{[{\bf A}_{nk}\sigma_{km} 
-\sigma_{nk}{\bf A}_{km}]}
\end{equation} 
Further, the identity 
\begin{equation}
\nabla_{\bf R}({\bf C}\cdot{\bf D}) = ({\bf C}\cdot\nabla){\bf D} + 
({\bf D}\cdot\nabla){\bf C} + {\bf C}\times(\nabla\times{\bf D}) 
 + {\bf D}\times(\nabla\times{\bf C})
\end{equation}
can be used to simplify the last term in Eq.(A6) to
\begin{equation}
\nabla_{\bf R}(\dot{\bf R}\cdot{\bf A}_{nm})=(\dot{\bf R}\cdot\nabla){\bf A}_{nm} -
\dot{\bf R}\times(\nabla\times{\bf A}_{nm})
\end{equation}
 Using this, the equation of motion for the nuclear coordinates becomes
\begin{eqnarray}
M\ddot{\bf R} = \sum_{m,n}{[\dot{\bf A}_{mn}\sigma_{nm} + {\bf A}_{mn}\dot{\sigma}_{nm}]}
               -\sum_{m}{\sigma_{mm}\nabla_{\bf R}E_{m}({\bf R})} -\nabla_{\bf R}U({\bf R})\nonumber\\ 
               -\sum_{m,n}{\sigma_{nm}[(\dot{\bf R}\cdot\nabla){\bf A}_{mn} + 
                \dot{\bf R}\times(\nabla\times{\bf A}_{nm})]}
\end{eqnarray}
  The time derivative of $\bf A$ can be rewritten as $\dot{\bf A}_{mn} = 
(\dot{\bf R}\cdot\nabla){\bf A}_{mn}$ and hence cancels with the corresponding term on the 
right hand side. Thus, the equation of motion simplifies to:
\begin{equation}
M\ddot{\bf R} = -\sum_{m}{\sigma_{mm}\nabla_{\bf R}E_{m}({\bf R})} -\nabla_{\bf R}U({\bf R})
                + \sum_{m,n}{[{\bf A}_{mn}\dot{\sigma}_{nm} - 
                  \sigma_{nm}\dot{\bf R}\times(\nabla\times{\bf A}_{nm})]}
\end{equation}
  Furthermore, the time derivative of the electronic density matrix, $\dot{\sigma}_{nm}$, can 
be replaced by using the Liouville-Von Neumann equations of motion for the electronic density 
matrix, to obtain
\begin{eqnarray}
M\ddot{\bf R} =  -\nabla_{\bf R}U({\bf R})-\sum_{m}{\sigma_{mm}\nabla_{\bf R}E_{m}({\bf R})}
                +i\sum_{m,n}{E_{mn}{\bf A}_{mn}\sigma_{nm}} - \nonumber\\
                 i\sum_{mn}{{\bf A}_{mn}\{\dot{\bf R}\cdot[\sigma,{\bf A}]_{nm}\}} -
                \sum_{mn}{\sigma_{nm}\dot{\bf R}\times(\nabla_{\bf R}\times{\bf A}_{mn})}
\end{eqnarray}
  The first three terms are the standard Ehrenfest force acting on the nuclear variables, 
while the last two terms are new. Further, using the vector identity
\begin{equation}
{\bf B}\times({\bf C}\times{\bf D}) = ({\bf B}\cdot{\bf D}){\bf C} - 
({\bf B}\cdot{\bf C}){\bf D} 
\end{equation}
The penultimate term on the right hand side can be simplified and rewritten as 
\begin{equation}
i{\bf A}_{mn}\{\dot{\bf R}\cdot[\sigma,{\bf A}]_{nm}\} = -i\sigma_{nm}\dot{\bf R}
\times({\bf A}\times{\bf A})_{mn}
\end{equation}
  putting this together with Eq.(A14) and defining an effective magnetic field by 
\begin{equation}
   {\bf B}_{mn} = \nabla_{\bf R}\times{\bf A}_{mn} - i({\bf A}\times{\bf A})_{mn}
\end{equation}
  the equation of motion for the nuclear coordinate is obtained as 
\begin{equation}
M\ddot{\bf R} = -\nabla_{\bf R}U({\bf R})-\sum_{m}{\sigma_{mm}\nabla_{\bf R}E_{m}({\bf R})}
                +i\sum_{m,n}{E_{mn}{\bf A}_{mn}\sigma_{nm}} - \dot{\bf R}\times
\sum_{m,n}{\sigma_{nm}{\bf B}_{mn}}
\end{equation}
  This is the equation of motion discussed in the text of the paper, and is the Ehrenfest
equation of motion with an additional contribution from the "magnetic field", $\bf B$. 
\nonumber


\newpage
\vskip1.0cm
\small{FIG.1: A schematic describing parallel transport of the electronic Hilbert space along a 
given nuclear path. The orthogonal axes represent the local electronic Hilbert space at a given nuclear 
position. As the nuclear path is traversed, the Hilbert space is transported along with it leading to the 
factor,$\Gamma$, which connects the Hilbert spaces at different nuclear positions ${\bf R}_{1},{\bf R}_{2}$
while $\bf R, Q$ are initial and final positions for the given nuclear path.} 
\par

\vskip1.0cm
\small{FIG.2: A pictorial representation of the nonadiabatic paths that contribute to the quantum 
dynamics. Two different paths corresponding to two nonadiabatic transitions are shown. The propagation along 
the horizontal lines is governed by the Born Oppenheimer propagator, while vertical columns represent 
nonadiabatic transitions. The widths, $t_{i}$, on the x-axis of the columns are the times over which  
nonadiabatic transitions take place. In the surface hopping approximation, these widths become 
infinitesimally small.} 

\newpage
\begin{figure}
\begin{center}
\vskip1.0cm
\includegraphics{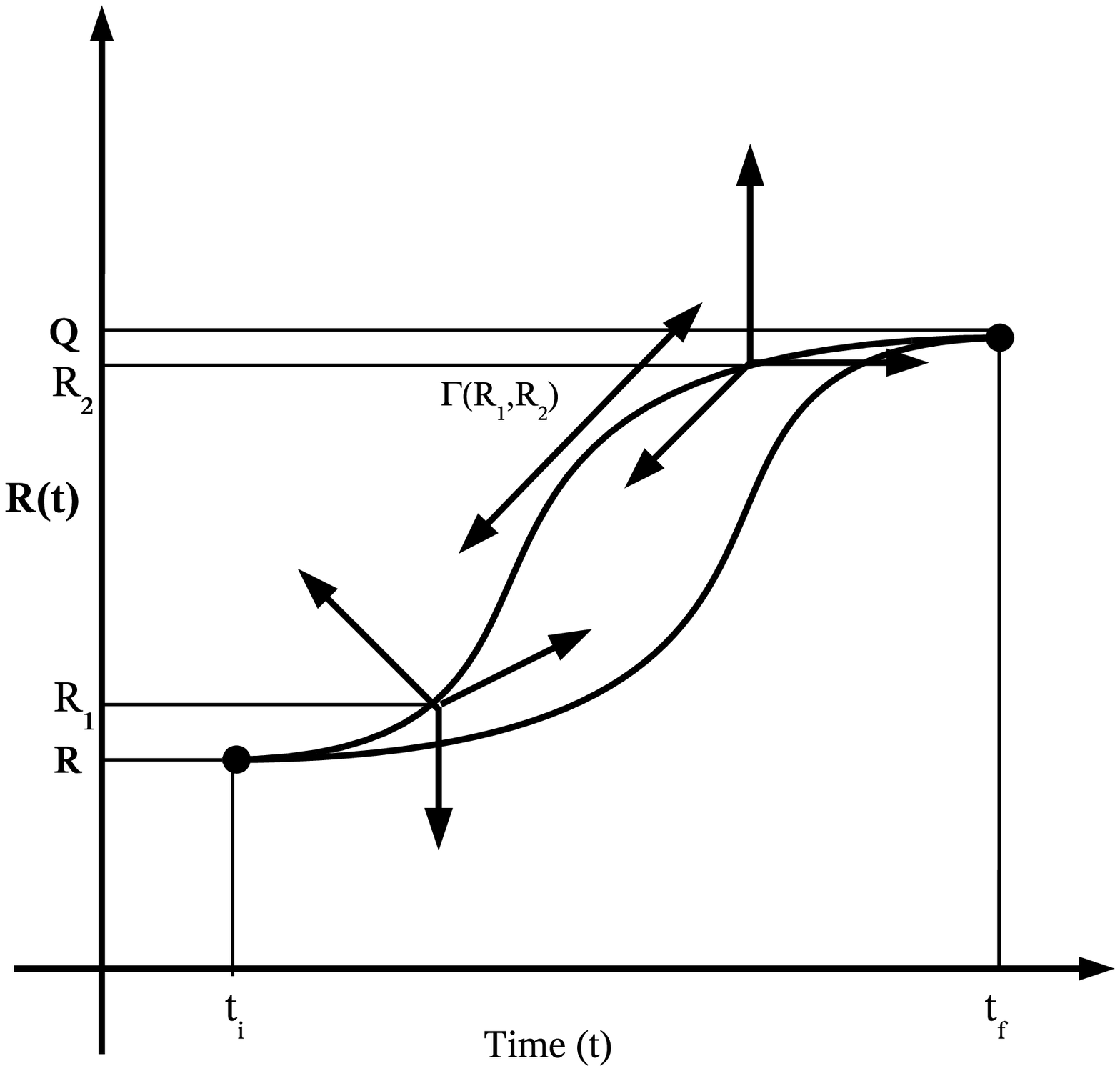}
\small{\caption{}}

\end{center}
\end{figure}
\newpage
\begin{figure}
\begin{center}
\vskip1.0cm
\includegraphics{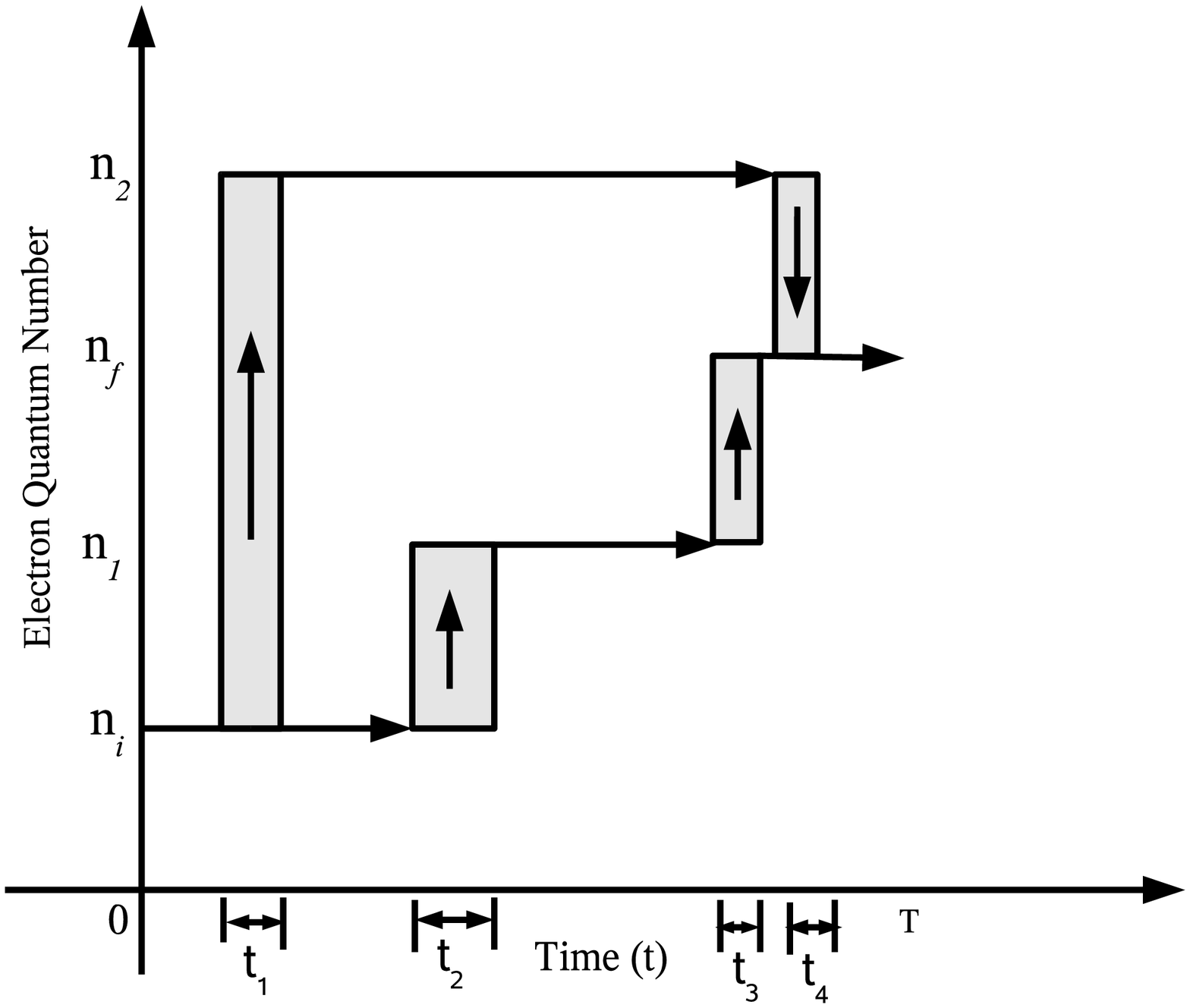}
\small{\caption{}}

\end{center}
\end{figure}

\end{document}